# Template Mediated Formation of Colloidal Two-Dimensional Tin Telluride Nanosheets and the Role of the Ligands

*Fagui He,[1] Eugen Klein,[1] Stephan Bartling,[2] Siavash Saeidpour,[3,4] Björn Corzilius,[3,4,2] Rostyslav Lesyuk,[1,5] Christian Klinke[1,4,6]\**

[1] Institute of Physics, University of Rostock, Albert-Einstein-Straße 23, 18059 Rostock, Germany

[2] Leibniz Institute for Catalysis (LIKAT), Albert-Einstein-Straße 29a, 18059 Rostock, Germany

[3] Institute of Chemistry, University of Rostock, Albert-Einstein-Straße 27, 18059 Rostock, Germany

[4] Department "Life, Light & Matter", University of Rostock, Albert-Einstein-Straße 25, 18059 Rostock, Germany

[5] Pidstryhach Institute for applied problems of mechanics and mathematics of NAS of Ukraine, Naukowa str. 3b, 79060 Lviv, Ukraine & Department of Photonics, Lviv Polytechnic National University, Bandery str. 12, 79000 Lviv, Ukraine

[6] Department of Chemistry, Swansea University – Singleton Park, Swansea SA2 8PP, United Kingdom

ABSTRACT We report the colloidal synthesis of 2D SnTe nanosheets through precursor hot-injection in a nonpolar solvent. During the reaction, an important intermediate – Sn-template – is formed which defines the confined growth of SnTe. This "flake-like" structure gives the first evidence for the possible 2D morphology formation prior to the anion precursor injection (TOP-Te). Additionally, we explore the role of each ligand in the reaction process. Thus, we explain the formation and morphology evolution of 2D SnTe nanostructures from a mechanism perspective as well as the role of each ligand on the molecular scale. The interplay of ligands provides the necessary conditions for the realization of stable low-dimensional SnTe nanomaterials with tunable size and shape.

**Introduction**

Low-dimensional thermoelectric materials such as tin telluride (SnTe) can adopt advantages based on the quantum confinement effect, suggesting great potential for heat-electricity conversion.[1] As a IV-VI narrow bandgap semiconductor (0.18 eV, bulk), SnTe exhibits an intrinsically high charge carrier concentration, which results in a relatively low Seebeck coefficient,[2, 3] but optimization of the material through doping and alloying offers great promise for thermoelectric applications of this material.[2] SnTe has also gained significant interest due to its exciting properties as a topological crystalline insulator, IR detection and radiation receivers material, as well as photovoltaic absorber.[4-8] So far, attempts to obtain solution-based SnTe nanocrystals (NCs) mainly yielded zero-dimensional (0D) or one-dimensional (1D) nanostructures.[9-11] Recently we reported a synthesis protocol for two-dimensional (2D) colloidal SnTe nanostructures.[12] Now, we disclose the important role of ligands in the colloidal synthesis process information of the 2D morphology and the observed faceting of nanocrystals. The obtained nanomaterials are investigated by means of scanning electron microscopy (SEM), transmission electron microscope (TEM), high-resolution TEM, powder X-ray diffraction (XRD), X-ray photoelectron spectroscopy (XPS), energy-dispersive X-ray (EDX) analysis and Fourier transform infrared (FT-IR) measurements.[13] Since SnTe has a cubic crystal structure with m$\bar{3}$m symmetry and octahedral coordination geometry, the formation of anisotropic shapes has to be conducted by molecular templates. These templates are well-organized domains consisting of the cation and a specific combination of ligands/counterions. When the tellurium precursor is injected into the reaction mixture, the tin/halogen arrangement of the template is replaced by the Sn-Te bond, and a bilayer oleic acid shell is formed at the same time. These findings show the importance of co-ligands and emphasize the formation of templates in the syntheses of colloidal nanomaterials.

**Methods**

*Chemicals and Materials:* Tin (II) acetate (Sn(CH$_3$CO$_2$)$_2$, anhydrous, ≥99.99%, stored in a nitrogen filled glovebox), tellurium shots (Te, amorphous, 1-2 mm; 99.999 %; stored in a nitrogen filled glovebox), oleic acid (90%), trio-*n*-octylphosphine (TOP; 97%; stored in a nitrogen filled glovebox), diphenyl ether (≥99%), and 1-chlorotetradecane (1-CTD, 98%) were purchased from Sigma-Aldrich. Ethanol and toluene were purchased from Honeywell. All the chemicals were used as-received without additional purification. Trio-n-ctylphosphine-Te (TOP-Te) was prepared and stored in a glovebox and all the syntheses were carried out applying standard air-free Schlenk-line techniques.

*Synthesis of 2D SnTe nanosheets:* In a three necked flask equipped with a condenser, a septum and a thermocouple in a glass mantle, 59.5 mg of Sn (CH$_3$CO$_2$)$_2$ (0.25 mmol), 2 mL of oleic acid (6.25 mmol), and 0.5 mL of tri-n-octylphosphine (1.1 mmol) were mixed in 10 mL of diphenyl ether, stirred at 130 °C for 5 min and degassed under vacuum at 80 °C for 1.5 h. Afterwards, the reaction solution was heated to reaction temperature 210°C under nitrogen after adding 0.15 mL of 1-chlorotetradecane (0.5 mmol). 10 min later, 0.8 mL of a 0.65 M tri-*n*-octylphosphine-tellurium (Te‑P(octyl)$_3$, TOP-Te) precursor solution was injected. The solution turned from clear yellow to greyish yellow. The reaction was quenched after 1 min by removal of the heating mantle. The resultant nanostructures were purified by precipitation with toluene, centrifugation at 4000 rpm for 3 min (3 times), removal of the supernatant and re-suspension in toluene for further characterization or storage.

*Synthesis of Sn templates:* In a typical synthesis of Sn template, 59.5 mg of Sn (CH$_3$CO$_2$)$_2$ (0.25 mmol), 2 mL of oleic acid (6.25 mmol), 0.5 mL of tri-n-octylphosphine (1.1 mmol) and 10 mL of diphenyl ether were mixed and dissolved in a 50 mL three-neck flask, stirred at 130 °C for 5 min and degassed under vacuum at 80 °C for 1.5 h. Afterwards, the reaction solution was heated to reaction temperature 210°C under nitrogen after adding 0.15 mL of 1-chlorotetradecane (0.5 mmol). 10 min later, the heating mantle was removed and when the temperature reached to 75°C, 15 mL of ethanol was injected into the reaction. The resultant white powder was then purified by centrifugation with ethanol at 6000 rpm for 3 min (3 times). The product could then be re-suspended in ethanol for further characterization or storage.

*Scanning electron microscope (SEM) and Energy Dispersive X-ray Spectroscopy (EDX):* Standard SEM images and EDX elemental mapping were performed on a *Zeiss EVO MA 10* microscope at an acceleration voltage of 10 kV. Samples were prepared by dropping 10 $\mu$L of the dilute NS solution onto a silicon wafer.

*Transmission Electron Microscopy (TEM):* Standard TEM images were performed on a *JEOL Jem-1011* microscope at an acceleration voltage of 100 kV. Samples were prepared by dropping 10 $\mu$L of the dilute NS solution onto carbon-covered copper grids.

*Powder X-ray Diffraction (XRD):* The crystal structure of the SnTe nanosheets was determined by XRD measurements, which were carried out on a *Philips X'Perts PRO MPD* diffractometer with monochromatic X-Ray radiation from a copper anode with a wavelength of 1.54 A (CuK**α**). The samples were prepared by dropping 10 $\mu$L of the dilute NS solution onto a silicon wafer.

*X-ray Photoelectron Spectroscopy (XPS):* measurements were performed on an ESCALAB 220iXL (Thermo Fisher Scientific) with monochromated Al Kα radiation (E = 1486.6 eV). Samples are prepared on a stainless-steel holder with conductive double-sided adhesive carbon tape. The electron binding energies were obtained with charge compensation using a flood electron source and referenced to the C 1s core level of carbon at 284.8 eV (C-C and C-H bonds). For quantitative analysis the peaks were deconvoluted with Gaussian-Lorentzian curves using the software Unifit 2021. A background model composed of a Shirley background + polynomial background was used and the parameters are varied during the fit together with the peak parameters to find the optimal background. The peak areas were normalized by the transmission function of the spectrometer and the element specific sensitivity factor of Scofield1.

*Sensitivity-enhanced magic-angle spinning nuclear magnetic resonance (MAS NMR) by dynamic nuclear polarization (DNP):* $^1$H-$^{31}$P CPMAS NMR measurements were performed by suspending the dried nanomaterials in a 1,1,2,2-tetrachloroethane solution containing 15 mM TEKPol polarizing agent. [37] Experiments were carried out using a Bruker Biospin ASCENT 400DNP wide bore magnet operating at 9.4 T (400 MHz $^1$H frequency) with a Bruker AVANCE III HD spectrometer and a Bruker 3.2 mm LTMAS DNP probe tuned to $^1$H and $^{31}$P in dual channel mode. Spectra were recorded at an MAS frequency of 15 kHz at a sample temperature of 100 K utilizing cross-polarization from hyperpolarized $^1$H under microwave irradiation from a Bruker/CPI 263 GHz gyrotron source operating at 130 mA beam current. The $^1$H-DNP enhancements for the frozen dispersions of SnTe-template and SnTe-nanosheets were 23 and 152, respectively.

*Fourier transform infrared (FTIR):* FTIR measurements were carried out by drying the nanomaterials and putting the powders on a diamond-ATR unit (PerkinElmer Lambda 1050+). The FTIR measurements are performed with a range from 500 to 4000 cm$^{-1}$.

**Results and Discussion**

The coordinating ligands can modify the surface energy of exposed crystallographic facets (which assume an increasingly prominent role with increasing surface-to-volume ratio) or binding selectively to particular facets, thereby favoring specific morphologies or yielding altogether different polymorphs.[14] In our previous work, the type and content of ligands used in the colloidal synthesis have been optimized; however, a fundamental investigation and understanding of the mechanism would be very beneficial for the precise control of the formation and faceting processes. In this report, first we used a large amount of ethanol to prevent the reaction before the hot injection of the tellurium precursor. As a result, a white powder, here called the Sn-template, was obtained. **Figure 1a** shows a SEM image of the Sn-templates. We observe a flake-like structure giving the first evidence for the possible 2D morphology formation prior to the tri-n-octylphosphine (TOP)–Te injection, which is similar to the Zn-soft template formation.[15] The selected area (red frame) in **Figure 1a** contains a large amount of Sn and carbon from the Sn-oleate complex, as well as a smaller amount of Cl which stems from the 1-chlorotetradecane (1-CTD) ligands (**Figure S1**). **Figures 1b** and **c** depict a TEM image and a XRD pattern of SnTe nanosheets (NSs) after synthesis and purification steps. The sample consist mainly of large nanosheets with lateral dimensions between 1 and 5 µm and a thickness of 57 nm calculated from XRD data using the Scherrer formula and 48 nm measured by atomic force microscopy (**Figure S2**). The difference in thickness stems from the three-dimensional cubic side products which contribute to the XRD signals and have edge lengths bigger than 100 nm. The XRD pattern shows only 2 signals due to a strong texture effect attributed to the (200) and (400) reflections. These signals can be attributed to the cubic crystal structure with the $Fm\bar{3}m$ space group.

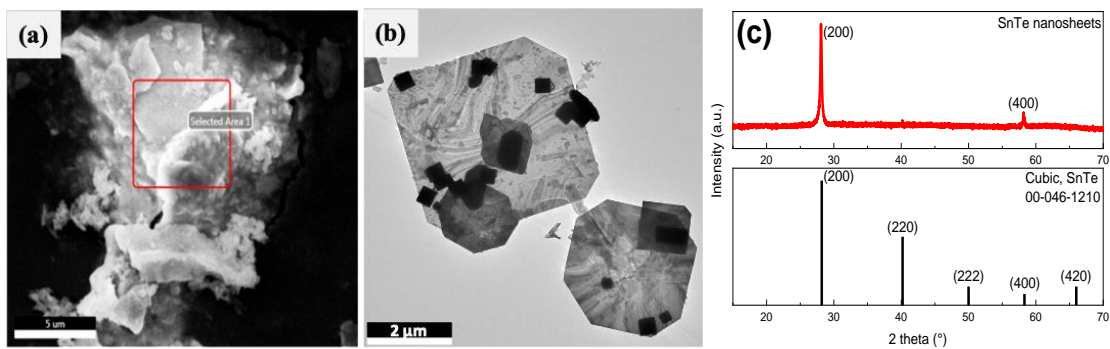

**Figure 1.** (a) SEM image of the Sn templates with flake-like structure; (b) TEM image of the product containing mainly large nanosheets and minor fraction cubic particles; (c) XRD pattern of the SnTe nanosheets with a strong texture effect.

In previous studies, we observed that halogenated compounds have an important influence on the shape of semiconductor nanomaterials.[12, 13, 16, 17] The active species promoting the shape transformation and confined growth were halide ions produced in situ which influence both the nucleation and ripening of the nanostructures.[13] We also evidenced that the amount of 1-chlorotetradecane (1-CTD) is one of the key factors for obtaining relative uniform and well-defined 2D SnTe nanostripes while only agglomerates were produced if no haloalkanes were used during the synthesis of SnTe nanostructures.[12] The importance of 1-bromoheptane (Br-Hep) in the formation of CdSe nanosheets was earlier revealed, suggesting that two active species, Br-Hep and ionic $Br^-$, were present on the surface of the CdSe nanosheets and partially replace the carboxylate ligands.[13] As the new surface ligands, the formation of Cd-Br bonds can regulate the growth rate by modulation of the relative surface energies of those facets and consequently influence the shape of CdSe nanosheets.[13] However, in the present work we find that 1-CTD influences the formation and final shape of SnTe nanosheets in a different way.

X-ray photoelectron spectra (XPS) were recorded to analyze the composition and the chemical states of all elements in Sn-templates and SnTe nanosheets. The survey spectra of both samples are shown in **Figure S3a** and **b**, respectively. As main elements C, O, and F as well as Sn, Te (for SnTe nanosheets), and Cl (for Sn-templates) with minor contributions can be identified. The XPS quantification data are shown in **Table S1.** High-resolution XPS scans of Sn 3d and Cl 2p of the Sn-templates, as well as Sn 3d and Te 3d of the SnTe nanosheets are presented in **Figure 2**. The Sn $3d_{5/2}$ peak of Sn-template can be found at a binding energy of 487.2 eV (**Figure 2a**), which can be assigned to $Sn^{2+}$. **Figure 2b** shows a rather small Cl 2p peak of Sn-templates at a binding energy of 199.5 eV which might be attributed to $SnCl_2$ as minor Sn species. We speculate that Cl does not exist as free $\overline{Cl}$ ions. This is in good accordance with the previous observations that $\overline{Cl}$ ions, which are introduced to the synthesis by chloroalkanes, are attached to the surface of the nanostructures in the form of a halogen–metal complex.[18] After the Te precursor (TOP-Te) was injected into the reaction, the product was further analyzed. **Figure 2c** shows the XPS spectrum of Sn 3d for SnTe nanosheets (see **Table S2** for details of the fit). The peaks at 487.0 and 495.5 eV correspond to the binding energies of Sn $3d_{5/2}$ and Sn $3d_{3/2}$ of SnTe ($Sn^{2+}$). However as shown for the template, the binding energy of $Sn^{2+}$ attributed to tin oleate is located nearly at the same spectral position. The peaks located at 488.8 and 497.3 eV indicate the oxidation state $Sn^{4+}$ observed in the $SnO_2$ crystals.[19] The weak peaks at 485.4 and 494.0 eV are assigned to elemental Sn. **Figure 2d** shows the XPS spectrum of Te 3d (see **Table S3** for details of the fit). The peaks at 574.4 and 584.9 eV are assigned to Te $3d_{5/2}$ and Te $3d_{3/2}$ in SnTe, respectively. The peaks at 572.4 and 582.9 eV could be assigned to the Te $3d_{5/2}$ and Te $3d_{3/2}$ from elemental Te and the weak peaks at 576.1 and 588.4 eV could be assigned to the oxidized species of $Te^{4+}$ in $TeO_2$.[20-22] The results express the clear propensity for oxidation of the SnTe nanostructures.[23] However, due to the formation of an oxide shell on SnTe nanocrystals, the synthesized material could be protected from being oxidized further and promote the stability of the nanocrystals.[11, 12] The data reveal also fluorine as very intense component of the surface, identified as contamination originated in the grease used for the glassware during the synthesis procedure. In fact, a majority of the F is bond to C as can be confirmed by the C 1s spectra shown in **Figure S4** with strong peaks at 291 and 293 eV corresponding to $C-F_2$ and $C-F_3$, respectively.

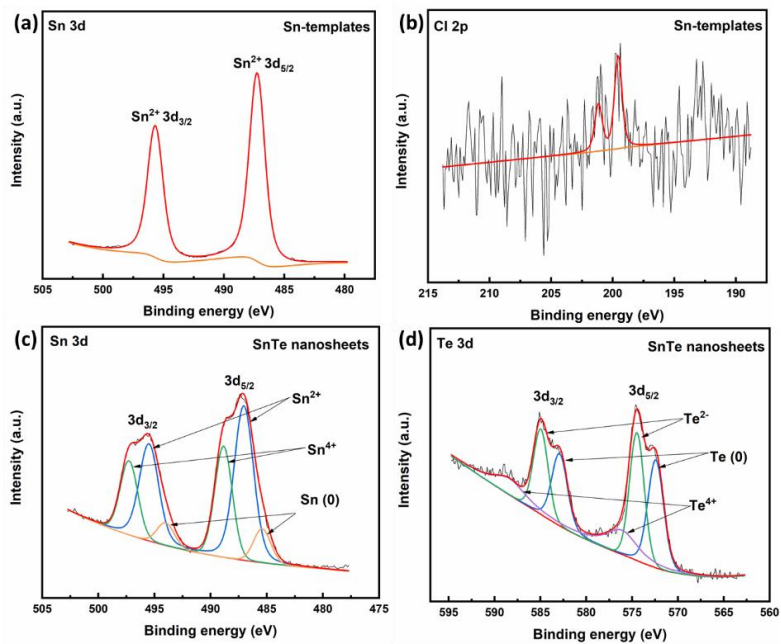

**Figure 2.** XPS spectra: Sn 3d (a) and Cl 2p (b) of Sn template; Sn 3d (c) and Te 3d (d) of SnTe nanosheets.

In the context of understanding the bonding mode of the ligands with the nanomaterial surfaces, FTIR spectroscopy has been employed. As seen in **Figure 3a** and **b**, the bands at 2853 and 2922 cm$^{-1}$ were attributed to the asymmetric CH$_2$ stretch and the symmetric CH$_2$ stretch modes of oleic acid, respectively.[24] It is worth noting that the band at 1710 cm$^{-1}$, corresponding to stretching vibration of C=O in pure oleic acid (see Supporting Information, **Figure S5**), was absent in the spectrum of Sn-templates. Instead, two new bands at 1457 and 1583 cm$^{-1}$ appeared in the FTIR spectrum of the Sn-templates (**Figure 3a**). They were attributed to the asymmetric (-COO$^-$) and symmetric (-COO$^-$) stretch vibration bands, respectively.[24, 25] This indicates that there is no free oleic acid in the Sn template sample and a complexation between the carboxyl and Sn was formed, which is in good accordance with the XPS results. As a result, oleic acid molecules are chemically adsorbed on the surface of Sn template through the chemical interaction between their -COO$^-$ groups and Sn atoms, meanwhile, the hydrophobic tails of oleic acid molecules face outwards (**Figure 3c**) and form a nonpolar shell which supports the single layer coated Sn-templates that can be dispersed in nonpolar carrier liquids. After TOP-Te was injected into the reaction, the band at 1710 cm$^{-1}$ appeared in the FTIR spectrum of SnTe nanosheets while the band at 1457 cm$^{-1}$, the asymmetric (-$\overline{COO}$) stretch vibration band, became very weak (**Figure 3b**). Thus, the FTIR signal changes also verify that the injection of TOP-Te to the synthesized oleic acid-coated Sn-template changes the kind of bonding between -$\overline{COO}$ groups of the oleic acid and the Sn atoms. The band at 1710 cm$^{-1}$ is the characteristic band of a secondary layer in bilayer oleic acid-coated nanomaterials which was already demonstrated by Wen et al.[24] The primary layer in the bilayer coated structure is chemically adsorbed on the surface of nanoparticles, and the secondary layer is physically adsorbed on the primary layer through the interpenetration of the tails of the primary and secondary surfactants at their interface (**Figure S6**).[24, 26] Because the oleic acid of the secondary layer was only physically adsorbed on the primary layer, the 1710 cm$^{-1}$ band due to the stretching vibration of C=O in oleic acid should appear in the FTIR spectrum of bilayer oleic acid-coated SnTe nanosheets.

The tri-*n*-octylphosphine (TOP)–Te complex was used as a tellurium precursor in the synthesis. It has a higher cleavage rate compared to TOP-S and TOP-Se, which leads to the formation of more nuclei and faster exhaustion of the monomers. This fast cleavage rate and a higher number of nuclei in the reaction showed an adverse effect on the anisotropic growth and resulted in 3D bulk structures instead of a 2D morphology.[27] Interestingly, the introduction of an appropriate amount of TOP is the guarantee for obtaining a two-dimensional structure along with the presence of the haloalkane. For comparison, reference syntheses with different amounts of TOP were also performed. No product could be obtained under these conditions without TOP while only agglomerates were produced if less or more amount of TOP compared to the standard procedure was injected into the synthesis (**Figure S7**). TOP plays an important role in slowing down particle growth, likely by blocking surface binding sites.[28, 29] When TOP was used as a capping ligand in the synthesis of nanomaterials, the coordination with the TOP ligands could effectively stop the growth of the nanoparticles (even in the presence of unreacted precursors) which would influence the final shape of nanomaterials.[30] In addition, TOP could act as a reducing agent as well, leading to producing SnO through an autocatalytic process at the SnTe NSs surface.[31]

Another significance of TOP that can ensure the acquisition of 2D structures is that TOP affects not only the reaction pathway for the halide transfer but also the dissolution of tin halides on SnTe nanocrystals which is similar to the structural development of CdSe tetrapods.[32] As shown in **Figure S8**, three resonances at ~-12, ~30 and ~50 ppm in the $^{31}$P NMR spectrum of both the Sn template and SnTe nanosheets indicate the occurrence of ligand exchange and elimination when halide and TOP are present in the solution: alkylphosphonium halides ($R_4P^+$(halide)$^-$ where R = alkyl chain), formed from the reaction between alkylhalides (thermal decomposition product) and neat TOP, and halide-bound Sn is dissolved by forming $(R_3P)_2Sn(halide)_2$.[32] When the binding between Sn and Cl was replaced by Sn–Te bonds, a certain quantity of Cl$^-$ ions is released into the solution, resulting in more ligand exchange between TOP and halide. Therefore, stronger resonance at around -12 ppm ($R_3P$) and 30 ppm ($R_4P^+$) are observed. On the other hand, in reactions where oleic acid was used as a surfactant molecule, the formation of II-VI nanocrystals was accompanied by the formation of trialkylphosphine oxides and $(OA)_2O$.[33] In particular, the conversion of TOP-Te to its corresponding phosphine oxides is linked to the formation of anhydrides of oleic acid, suggesting that phosphonic and carboxylic acids are responsible for the cleavage of the phosphorus-chalcogen double bond.[33] Thus, after the hot injection of TOP-Te, stronger resonance at around 50 ppm (corresponding to the conversion of TOP-Te to TOPO-Te) is observed due to the cleavage of TOP-Te. As H-OA is responsible for the cleavage of the P=Te bond, changing the concentration of these surfactants will likely change the TOP-Te cleavage kinetics in addition to the binding of surfactants to the nanocrystal surface. This is especially important because the rate of this cleavage will influence particle nucleation and growth and offers an explanation of the influence of the amount of OA on the morphology of the ultimate product (**Figure S9**).

Additionally, we investigated the influence of the other ligands and the synthesis temperature. The amount of oleic acid varies the lateral dimensions. Low amounts of oleic acid yield small nanosheets while larger amounts caused big nanosheets (**Figure S9**). Since the Sn templates consist mainly of tin cations and a combination of oleate and oleic acid molecules, a higher amount of oleic acid means larger and better-organized templates. On the other hand, the co-ligand 1-CTD also influences the synthesis. After a cleavage of the $Cl^-$ ion from the alkane rest, these small ions replace some of the oleate ligands in the template structure which alters the reactivity of the templates. An excessive amount of haloalkane molecules yield solely three dimensional (3D) structures while low amounts favor a large shape and size distribution (**Figure S10**). It is important to note that the cleavage of this molecule does depend on the temperature and the time it spends in the reaction mixture, which means that the same amount of 1-CTD could produce different products when changing the reaction conditions. The temperature mainly determines the reactivity in a direct and indirect way as mentioned before through the cleavage rate of the 1-CTD. When the reaction temperature is below 170 °C no reaction takes place for this material. Low reaction temperatures like 170 °C or 180 °C yield mainly 3D structures (**Figure S11**) while high temperatures like 250 °C produce products with a high size and shape distribution.[12] The results indicate that temperatures below 190 °C are not enough to separate the $Cl^-$ ions from the alkane rest, which means that the synthesis is basically performed without this co-ligand and yields solely 3D structures. At high temperatures, the cleavage rate of the 1-CTD is very high. Thus, the structure of the templates becomes more unstable resulting in templates with different shapes and reactivities.

After hot injection of the TOP-Te precursor, the colorless solution transiently turned to yellow (the color of TOP-Te precursor) and the morphology evolution was investigated by taking aliquots during the reaction followed by TEM analysis. At the early stage, when the color of the solution starts to change (from yellow to dark yellow at 32 s after hot injection of TOP-Te), the SnTe nanoplatelets with hexagonal shape (≈2 μm length, ≈400 nm width) could be observed (**Figure S12a**, Supporting Information). As the reaction proceeds, the nanoplatelets grow longer and the widths become larger (≈3 μm length, 600-700 nm width) (**Figure S12b**). Based on evidence from XPS and FTIR measurements shown above, we propose a possible formation process of SnTe nanosheets. As depicted in **Figure 3d**, the two ligands of oleic acid and 1-CTD help Sn to form a flat structure, the Sn template. After introduction of the Te precursor, the $Cl^-$ ions are replaced by Te to form Sn-Te bonds. Meanwhile, the SnTe nanostructure is coated with a well-organized primary oleic acid molecule. Then, the excess oleic acid was weakly adsorbed on the primary layer of the oleate-coated SnTe nanosheets to form a double layer shell through the steric intermolecular interaction between the subsequent molecule and the hydrophobic tail of oleate of the primary layer.

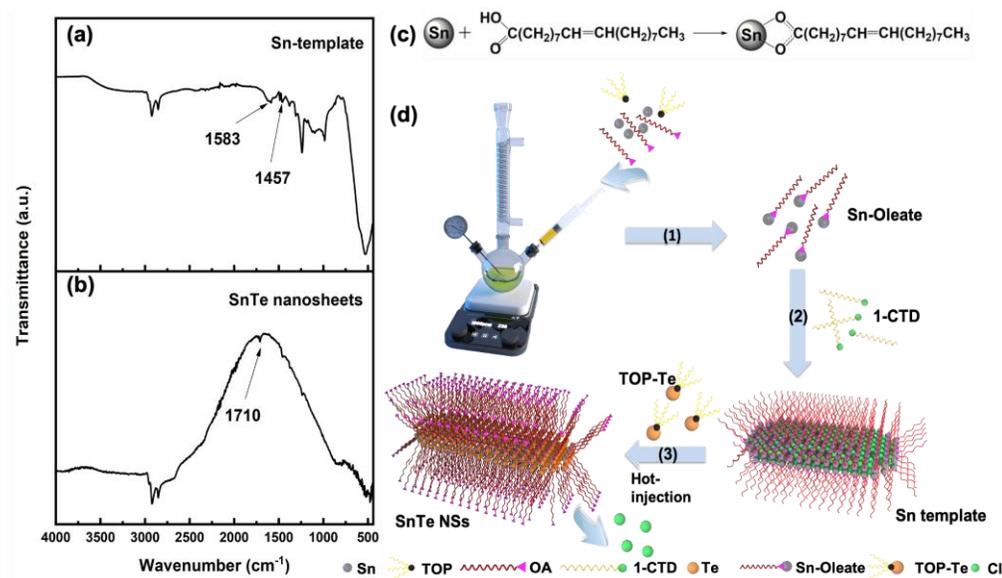

**Figure 3.** FTIR spectrum of Sn template (a); FTIR spectrum of final product SnTe nanosheets (b); scheme of the formation of Sn oleate (c); scheme of the proposed mechanism of the formation of SnTe nanosheets (d).

**Conclusion**

In this work, we discussed the interaction of ligands with metal precursors at different stages of the synthesis and investigated the formation process of SnTe nanosheets. We see from the FTIR analysis that the formation of an OA-double layer is already evidenced at the stage of the Sn-white-complex (template). At the same time, the Cl-interaction with Sn is evidenced by XPS, suggesting that both OA and halogen ensure conditions for the 2D templating. Later on, Cl disappears and only OA and TOP stay on the surface leading to growth in three dimensions with different velocities. Thus, the collective interplay of different ligation agents is needed for obtaining the 2D morphology in the cubic system of SnTe. Based on the experiments, we explained the formation and morphology evolution of 2D SnTe nanostructure from a mechanism perspective as well as the role of each ligand at the molecular scale which provides the building block for the realization of stable low-dimensional SnTe nanomaterials with tunable size and shape.

## ASSOCIATED CONTENT

**Supporting Information.**

EDS analysis of Sn templates; AFM images and measured height images of synthesized SnTe nanosheets; XPS spectra and XPS quantification data; fit parameters of the Sn 3d spectra and Te 3d spectra; FT-IR spectrum of pure oleic acid; TEM images; $^{31}$P NMR spectrum (PDF)

## AUTHOR INFORMATION

**Corresponding Author**

* E-mail: christian.klinke@uni-rostock.de

**Funding Sources**

This work was supported by the China Scholarship Council and German Academic Exchange Service (DAAD). Funded by the Deutsche Forschungsgemeinschaft (DFG, German Research Foundation) – SFB 1477 "Light-Matter Interactions at Interfaces", project number 441234705.

**Notes**

The authors declare no competing financial interest.

ACKNOWLEDGMENT

We would like to thank Fabian Strunk for the help with the XPS measurements. We acknowledge the European Regional Development Fund of the European Union for funding the PL spectrometer (GHS-20-0035/P000376218) and X-ray diffractometer (GHS-20-0036/P000379642) and the Deutsche Forschungsgemeinschaft (DFG) for funding an electron microscope Jeol NeoARM TEM (INST 264/161-1 FUGG) and an electron microscope ThermoFisher Talos L120C (INST 264/188-1 FUGG) and for supporting the collaborative research center LiMatI (SFB 1477).